 \documentstyle[epsfig]{article}
\newcommand{\cl}[1]{\begin{center} {#1} \end{center}}

\begin{document}
\rightline{SBNCBS/NUP/96/01}
\vskip 1.0truecm

\cl{\Large{Fluctuation dynamics and formation of exotic shape}}

\vskip 1truecm
\cl{     S. Chattopadhyay\footnote{e-mail~: shila@tnp.saha.ernet.in} }
\vskip 1truecm

\cl{ S.N. Bose  National Centre for Basic Sciences, Block-JD, Sector 3,
Salt Lake, Calcutta-700091, India}
\vskip 2truecm

\begin{abstract}
To study the role of fluctuation in the collisional dynamics, Boltzmann- Langevin 
formalism is applied in a two dimensional scenario. The importance of collective 
flow towards the formation of hollow structure is exhibited  in our simulation 
result. The observation of large clusters along the transverse direction 
indicates the non-equilibrium aspects of the fragmentation process in the 
late stage of the reaction. 
\end{abstract}


 \newpage
Numerical simulations based on  Boltzmann-\"Uhling-Uhlenbeck (BUU) 
equation have been carried out extensively 
for intermediate energy heavy ion collision. These models have their origin 
in effective one body theories, happens to be very useful in understanding 
a wealth of nuclear properties \cite{Bert}. The standard treatment (usually done in 
the N\"ordheim approach) takes into account only the average effect of the 
collision between the particles of the considered system. However, large 
scale fluctuation in density that manifests itself as fragments observed 
in the experiments cannot be described 
properly within the framework of single trajectory evolution of nuclear 
density. It is to be mentioned that owing to the presence of numerical 
noise  one actually observes fluctuation 
in the solution of the BUU equation. This feature becomes very crucial 
when the di-nuclear complex passes the point of instability where any 
fluctuation whatsoever small  can grow indefinitely. As a result 
the system may take any unphysical path in the  trajectory variable space. 

 In this context, the stochastic transport models provide a more 
appropriate starting point. In the intermediate energy regime, noise 
associated with the two-body collision term is considered to be the dominant 
source of fluctuation. To retain this effect in the one body level within 
the framework of BUU equation different models \cite{Ayik} and several computational 
schemes following from these models have been proposed [3 - 6].  Accordingly, the one-body 
distribution function $f({\bf r},{\bf p})$ within an elementary cell around 
a  point (${\bf r},{\bf p}$)  of the related phase space 
of the system is considered to be the trajectory variable. A point along 
the trajectory is specified by the values of each $f({\bf r},{\bf p})$ 
over a given space of $({\bf r},{\bf p})$ at time t. To 
study the fluctuation around the deterministic {\it i.e.} the mean 
trajectory, the  usual method is to incorporate fluctuating collision term in 
the BUU equation.  The resultant equation 
for the evolution of phase space density is commonly known as Boltzmann- Langevin 
(BL) equation. Strictly speaking, for fermionic dynamics the value of the 
phase cell occupancy factor $n_i$ (which  differs from $f({\bf r},{\bf p})$ by a 
constant multiplicating factor $\Delta{\bf r}\Delta{\bf p}/h^D$, the phase 
cell volume in D dimensional physical space) takes a value 
either 1 or 0 . This feature automatically fulfills the relevant statistical 
 relation related to the sample mean $\langle n\rangle$ and its variance  
$\sigma_i=\langle n\rangle_i \langle\bar n \rangle_i$. 
It is to be noted, however, that this feature does not allow us to make an 
attempt to produce correct 
magnitude of the noise by arbitrarily changing the phase 
cell occupancy within the 
samples from the known single or mean trajectory result calculated beforehand. 
In such cases, conservation principles related to the particle number, momentum and 
energy  cannot be satisfied for every individual 
sample and dynamical correlation among the phase cell occupancies 
is completely lost.
Therefore, the only possible option is to consider the evolution 
of each of the samples separately. Obeying the above 
mentioned criteria a novel simulation scheme is put 
forward for dynamical allocation of phase cell occupancy \cite{Chat}. 
The simulation result reproduces the correct average behavior 
as compared with the mean trajectory result (in other words the solution 
of the BUU equation) for a test case where both of the results should be identical.

 Most of the investigations so far done 
on the basis of BL model are mainly directed to study 
the onset of clusterization in the nuclear matter situated 
inside the spinodal zone \cite{Burg1} which clearly shows the inadequacy of mean trajectory 
description of such catastrophic phenomenon. 
For this particular case, appropriate theoretical 
analysis has also been done for studying the  growth of density fluctuation  
\cite{Colo1,Chat1}  
which in turn, provides an opportunity to check the reliability of the 
different simulation schemes. In this situation, fragments may appear 
anywhere within the volume with no significant spatial correlation. 
However, for a finite nuclear system the 
process of fragment formation may take place in a different manner.  
The standard mean trajectory calculation near Fermi energy domain 
explores the other alternatives \cite{More}. Within a specific incident 
energy  window of 50-100 MeV, the di-nuclear system takes exotic shapes 
due to the presence of instability and 
the fragments appear essentially by breaking of such structures. Later 
works have confirmed in a more rigorous manner such possibilities \cite{Xu} even 
at lower collision energies \cite{Bord}. 
This interesting observation 
promptly initiated some theoretical calculation to find out the experimental 
signatures of such facts \cite{Phai}, but there is still no conclusive 
experimental evidence
\cite{Brun}. From the theoretical point of view, when the issue is to study the 
dynamical evolution 
of the system in the presence of well developed instability, it seems more 
appropriate to probe the associated trajectory variable space around the mean 
path. This can be realized within the BL model  by 
allowing the evolution of a given number of samples individually. Fluctuation 
appeared in this theory as a dynamical variable rather than as a numerical 
noise. This method thus serves to test the extent to which the mean trajectory 
calculation is reliable. With this notion, in the present work we 
study the dynamical evolution of the collision process in the trajectory 
variable space. However, 
our study is confined to the 2D physical space where 
some of the essential features of the dynamics of fragment formation can still 
be unraveled. 

In the present study we consider the head-on collision between two identical 
spherical blobs in 2 dimension, each of them having rms radius of 5 fm. 
which is the same as that of the nucleus of mass number $\approx$ 100.  For 
the effective one body field we employ a simplified Skyrme interaction \cite{Colo2,Chat1}
\begin{equation}
U(x,y)=A{\rho(x,y)\over \rho_0}+B\Big({\rho(x,y)\over\rho_0}\Big)^2
\end{equation}
with A=-100.2 and B=48 which correspond to the value of compressibility modulus 
$K=288$ MeV  
and  Coulomb interaction is included for protons. 
To evaluate effective potential on the co-ordinate grid, an averaging over 
2D gaussian function of width 0.87 fm is considered. For the simulation  
on phase space lattice we use (27$\times$ 27) cells in momentum space having width 
$\Delta p\approx$ 58 MeV/c which covers the range $\pm$ 780 MeV/c in each 
of its direction and the corresponding cell width in co-ordinate space $\Delta x$ is 
taken to be ${1\over 3}$ fm. Each filled phase cell can be considered as a 
particle. To describe nuclear matter at saturation density 
$\rho_0$ (=0.55 fm$^{-2}$) in a unit coordinate cell one needs $2\times 69$ 
such particles such that a single nucleon is represented by nearly 2200 
particle states for every sample. For an average description of the 
evolution we consider simulation of 20 such identical samples separately. 
In order to express the total energy of the di-nuclear system in the initial 
state as the sum of binding energies and the energy related to the boost, the 
boost momentum should be chosen as a simple multiple of the grid width 
$\Delta p_x$ only (say $B_{fac}$). This will also yield an accurate 
determination of the component of the flow velocity ${\bf u}(x,y)$   
at every spatial cell initially and in subsequent time . 
In the present work the values of $B_{fac}$  are taken to be 2, 3 and 4  
which correspond to the values of incident energy $E_{lab}$
32, 68 and 117 MeV/nucleon, respectively. 

The solution for the Vlasov part of the BUU equation
is realized within our model by standard Leap-Frog routine similar to that 
used in `test-particle' approach. (see ref.\cite{Chat1}, for details). It is to be noted 
that due to the presence of fluctuating collision term, the symmetry in 
the coordinate as well as in the momentum space of the initial distribution is 
gradually lost. This feature may triggers a certain situation so that in the 
process of Vlasov propagation  more than one particle may try to access  
the same phase cell. This disturbing feature is reduced substantially by simply 
choosing the time step of the simulation $\Delta t=m{\Delta_x\over\Delta_p}$ 
($\approx 5$fm/c in our case) which is 
exactly the time needed for a particle to move at least one single step in 
the spatial grid. It is observed that in this collisional scenario the number 
of such events always happens to be
less than  even 10 for all values of $B_{fac}$ mentioned above, whereas the 
total number of successful collisions may reach  a value more than 
40,000 in a single time step for an individual sample. 

Let us concentrate on the simulation result of the collision 
between two identical blobs in 2D.  We mainly discuss here 
the case related to the value of $E_{lab} = 68$ MeV/nucleon. The sample 
average density distribution $\rho(x,y)$ in units of $\rho_0$ is plotted 
in Fig. 1 for different stages of the collision. It is 
observed that within the time 30 fm/c the di-nuclear complex attains 
maximum compression. The central density reaches the value $\approx 1.6 \rho_0$. 
After that, the matter flows radially outwards where the flow velocity along 
the transverse direction of the collision i.e. along y axis, is much higher 
than that along x axis. An interesting feature is observed
during decompression at a time $t \approx 60$ fm/c. 
A narrow circular band  appears  near the periphery 
of the composite system (see Fig. 1(a)) and later on, matter deposits continuously 
around this band which makes it wider due to the outflow of the matter
from the central region. This process continues until the central density 
drops down to a very low value which is even less than 5$\%$ resulting in the 
formation of a ring-like structure. However, the matter distribution along the 
ring is not uniform. Relatively stronger flow along y axis  
even at very late stage of the collision produces eventually 
much larger deposition of matter near y axis. 
The maximum density here 
attains a value of about 0.60$\rho_0$ at $t = 200$ fm/c, and in  other parts 
of the ring the average density takes the value  $\approx 0.30\rho_0$. 
It is to be mentioned that the stability 
of this exotic shape depends crucially on the magnitude of the flow velocity. 
At a lower collision energy of $E_{lab}=32$ MeV/nucleon, 
the structure  does not expand much due to smaller values of flow velocities. 
Although we observe central 
rarefaction in the matter density which, however, is smeared out  
completely around the time 250 fm/c and a much  elongated residue having 
uniform density of $0.8 \rho_0$ is left. 

From this study it is revealed that the appearance of the small band 
like structure or a notch in the density distribution at the earlier stage of the 
collision is the key feature which is instrumental in the formation of exotic shape. 
In order to gain additional insight to its formation, we compute the  
diagonal components of the pressure 
tensor ${\bf P}$, the spatial gradiant of which  is essentially connected to the 
temporal variation of the flow velocity ${\bf u}(x,y,t)$. In Fig. 2 
we plot the components 
$\langle P_{xx}\rangle$ and $\langle P_{yy}\rangle$ in the rest frame of the 
fluid  along y axis which are 
computed by the appropriate  relation as given below 
\begin{eqnarray}
P_{\alpha\beta}(x,y;t)&=& \int (p_\alpha-\bar p_\alpha)(p_\beta-\bar p_\beta)
f(x,y,p_x,p_y;t)d{\bf p}
\nonumber\\
&+&\delta_{\alpha\beta}\Bigg({A\over 2}{\rho(x,y;t)\over \rho_0}+{2\over 3}B\bigg({\rho(x,y;t)\over 
\rho_0}\bigg)^2\Bigg)\rho(x,y;t)
\end{eqnarray}
where  $\bar p_\alpha=m u_\alpha$.
As the reaction proceeds  the pressure which 
initially is zero everywhere increases 
towards the center with the gradual piling up of matter there. Near 
the boundary, on the other hand, the low value of density may lead to a negative  
pressure. In general, 
the location of the minima  of the pressure profile depends on the 
particular choice of the equation of state one uses for simulation. 
It is to be noted that due to the continuous 
outflow of matter and moreover, since pressure is zero at the vacuum, these points 
of minimum 
stay near the matter boundary until the density at the  central region  drops 
down substantially. 
Therefore, a pocket is formed  as shown in the Fig. 2(a) where 
matter can accumulate  or be trapped so that  a band like structure  appears 
in the density profile. with time this pocket moves outwards as the system expands. 
On the contrary, no such pocket is observed in the profile of 
$P_{xx}$ along y direction (see Figs. 2(a) and 2(b)). 
The importance of asymmetry of the momentum distribution 
on the formation of this band may be understood from  the differences in 
magnitude of $P_{xx}$ and $P_{yy}$. 
To elaborate it further, we plot the the magnitude of 
$\langle Q(x,y)\rangle$ along y axis in Fig. 3 in accordance with the following 
definition 
\begin{equation}
Q(x,y)={{\int\bigg((p_x-\bar p_x)^2-(p_y-\bar p_y)^2\bigg)f(x,y,p_x,p_y)
d{\bf p}}\over{\int
\bigg((p_x-\bar p_x)^2+(p_y-\bar p_y)^2)\bigg)f(x,y,p_x,p_y)d{\bf p}}} ~.
\end{equation} ~.
The magnitude of the flow velocity $\langle u_y(0,y)\rangle$ is also 
depicted in the same graph. The x component of the velocity $u_x$ along y axis 
is found to be almost zero through out the evolution. 
The deposition of matter near the boundary means 
a negative slope to the  flow velocity ${\bf u}({\bf R})$ at these points which 
otherwise, is expected  to be a rising function of the radius  $R$ \cite{Dan}. 
The total landscape of $\langle Q\rangle(x,y)$  can be constructed from similar parallel 
curves  as shown in Fig.3(a) and 3(b). Simultaneous 
attainment of high values of both the quantities $\langle u_y\rangle$ and 
$\langle Q\rangle$  imples the presence of collective flow (Figs. 3(a) and 3(b)).  
On the other hand, if such a high value of $u_y$ is attained due to the  
collisions among particles only, then one should expect an opposite trend, 
{\it i.e.} the value of $u_y$ then increases with 
decreasing value of $Q$.  
In spite of the fact that the matter flows outwards with very high velocities and 
moreover, 
near the boundary, its magnitude exceeds that of the local sound velocity, 
a rarefaction wave behind this shock proceeds towards the center. 
This feature is attributed to a certain amount of fluctuation  
in the spatial variation  of $P_{yy}$ (see Fig. 3(c)). As a result, we observe 
a plateau 
in the landscape of flow velocity  which in turn produces depletion 
of matter behind the shock. From the above analysis it appears 
that strong collective flow and also a specific 
behavior of $Q$ that arises in the earlier stages of the collision, happens 
to be quite important for the formation of the hollow structure. 
We  observe that equilibration attains earlier in the 
peripheral zone of the residue as seen from Fig. 3(d) and is also evident from 
Fig. 2(d). Due to this typical behavior of $Q(x,y)$ 
the process of the formation of hollow structure is now being hindered, which is 
also apparent from a closer look at the 
the spatial variation of $P_{xx}$ and $P_{yy}$ in Figs. 2(d) and 
2(e). 
This is probably the real cause why in low energies the structure ultimately 
collapses towards the center and the role of external force as produced by 
Coulomb interaction becomes important in later stages. 
Eventually, when the central density drops down substantially with respect 
to that 
near periphery so that the pressure at center becomes less negative,  
the process of the formation of structure becomes self-sustained. This feature 
is essentially connected to the very nature of the nuclear matter below subnormal 
densities as has already been pointed out in Ref. \cite{Dan} and  observed in our 
simulation (see Fig. 2(f)). Along the beam direction {\it i.e.} 
along the x axis, one  observes a `stopping'  as the central density 
reaches its maximum value.  
However, at large x, matter still flows towards the center. At later times, due 
to the expansion, matter flows radially outward. 
The magnitude of the resultant of two opposing flow velocities along x axis 
is less than that along y direction. 
As a result, in course of time, the residue gets an elongated shape with 
larger deposition of matter along the transverse direction. 

Another important aspect of the dynamics, the process of clusterization, can be 
accommodated in a proper manner within the present formalism. Although we 
observe considerable smoothness in sample averaged density, however, in the 
later stage of the collision large scale fluctuation in the density is present
in an individual sample.  After a time of $\sim 100$ fm/c when the overall density 
of the residue decreases down to a value of $\approx 0.4 \rho_0$, 
the fluctuation becomes appreciable indicating the onset of clusterization. 
It is observed 
that $\delta\rho$($=\sqrt{\sum(\rho_i-\langle\rho\rangle)^2}$) is nearly 
proportional to the density $\rho$ itself. Therefore, the landscape of $\delta\rho(x,y)$ 
looks similar to that of $\rho$ as shown in Figs. 1(b) and 1(c). However, 
the proportionality 
constant of $\delta\rho/\rho$ gradually increases with time and attains a 
value $\sim 0.6$ at a time t=200 fm/c. The wavelengths of the dominant 
unstable modes corresponding to the observed density fluctuation \cite{Burg,Ayik1}
are much larger than the width of the ring formed as shown 
in Fig. 1(c). Therefore the finite size effect due to the relatively 
smaller width of the ring limits  
the role of spinodal instability in the fragment formation process. 
On the other hand, due to the continuous stretching along both  y and x 
direction, the structure breaks-up into several fragments.  
This feature is exhibited in Fig. 4. 
Two large clusters along y direction are always present in each of the samples 
and 
along with it several small size clusters and elongated pre-clusters 
also appear like beads around the rim of the elongated ring. 
Similar features have been also observed at a higher incident 
energy of $E_{lab} = 117$ MeV/A. In this case, the average density at the final 
stage is quite low $\approx 0.20\rho_0$ and the density along y 
direction near the periphery is still higher 
attaining a value of $\approx 0.4\rho_0$. It is to be noted that due to the faster 
rate of stretching here, arising from the relatively larger collective flow, 
the size of the clusters becomes 
very small which disappear gradually from the spatial grid. However, 
to check evaporation of matter from the surface of small clusters, the time step of the 
simulation at this stage should be decreased further. A better option is to 
use the standard Vlasov algorithm in the continuous space rather than in the 
lattice space so that time step can be chosen as small as required. 

In conclusion, within the BL formalism a simulation study has been performed 
to investigate the dynamics of fragmentation process. In the less complicated 
scenario in two dimensions, we analyze the role of 
collective flow towards the formation 
of exotic shapes. It drives the system in such a manner so that the composite 
nuclear object along the trajectory can reach the saddle of the deformation energy 
and subsequently the system appears in a hollow shape. 
Because of the fact, that Pauli blocking effect is included 
properly within our simulation method and moreover, the representative 
particles obey proper statistical criteria, the freeze-out temperature of 
the residue can be determined accurately from sample average values of kinetic 
energy and number of particles at particlular spatial grid. However, the 
observed fluctuation  in $\rho$ is larger than that associated to the 
temperature extracted from the average description.   
It is also interesting to note, that in spite of the presence of substantial 
amount of fluctuation in density at later stages of the collision, 
the related dynamics of the formation of structure is mainly controlled 
by the average description of the evolution. Therefore, we are still in the 
linear regime of the fluctuation dynamics. The history of the formation 
of structure or in other words, the entrance 
channel effect even persists in the process of fragmentation which is reflected 
by the production of large clusters along transverse direction. 
A proper 3D calculation along this line is called for  
to test the predictive power of this simulation scheme. By this means, it 
may be possible  to assess the reliability of  
mean-field models based on semi-classical approaches 
for the description of nuclear fragmentation.
 

%
%
%
%
%
 \newpage
 
\begin{figure}
\begin{center}
\epsfig{file=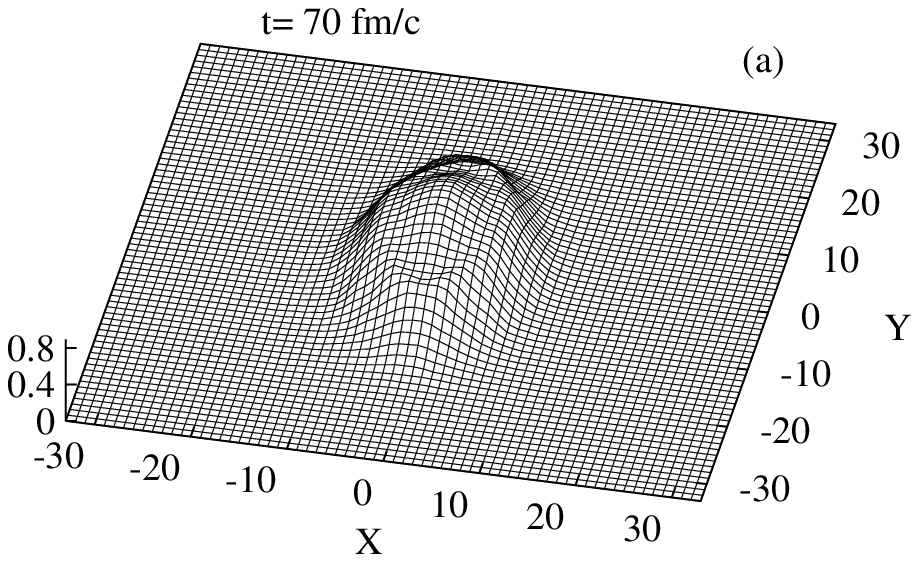,width=1.0\textwidth}
 \vspace{-1.3cm}
\epsfig{file=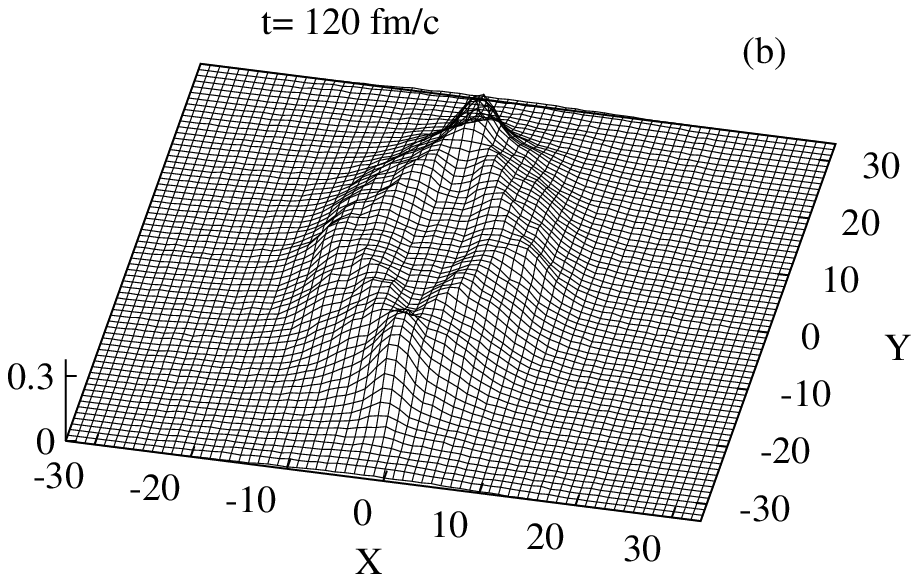,width=1.0\textwidth}
\end{center}
\end{figure}
\newpage
\begin{figure}
\begin{center}
\epsfig{file=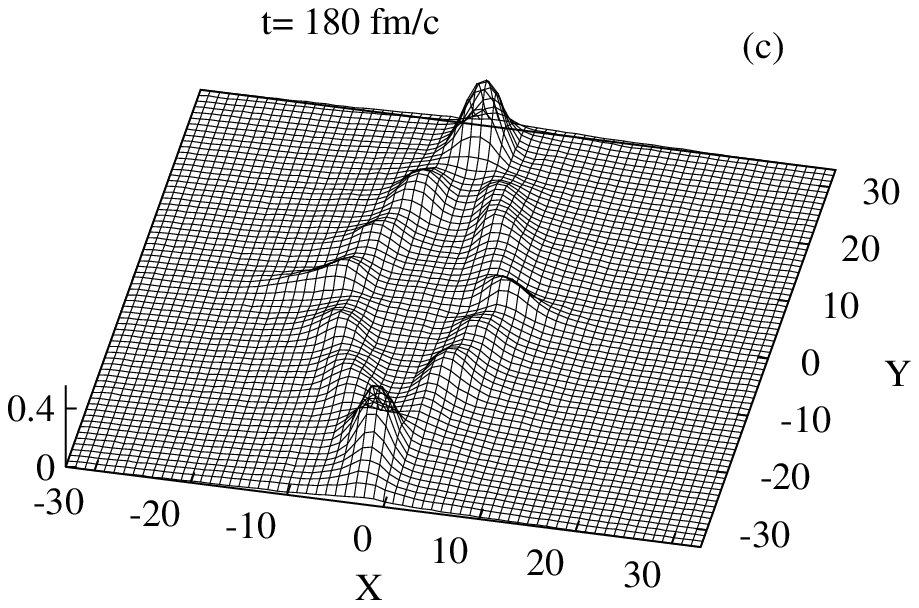,width=1.0\textwidth}
 \vspace{0.5cm}
\end{center}

{\bf Fig.1} \\
The sample average density profile $\langle\rho(x,y;t)\rangle/\rho_0$ versus 
the position ($x$, $y$) is shown for different times. 
\end{figure}

\newpage
\begin{figure}
\begin{center}
\epsfig{file=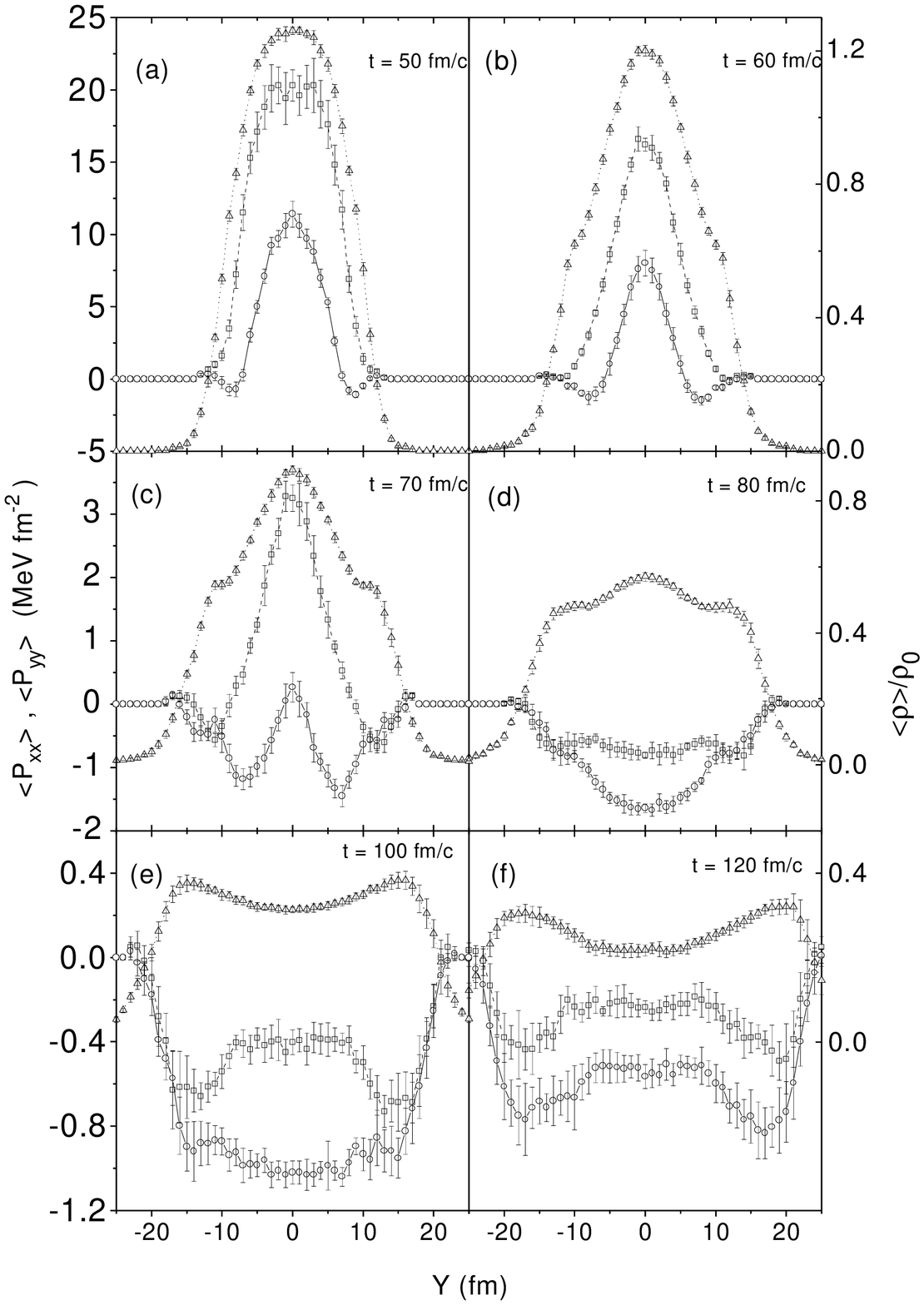,width=0.8\textwidth}
 \vspace{-1.0cm}
\end{center}

 {\bf Fig. 2} \\
The sample averaged values of the diagonal components of 
pressure tensor, $\langle P_{xx}\rangle$ and $\langle P_{yy}\rangle$ along 
y axis are shown by squares and circles, respectively for different 
time steps indicated in the figure. The sample averaged value of density 
$\langle\rho\rangle$ in units of $\rho_0$ along 
the y axis also shown by triangles in the same graph. The sample 
fluctuations are indicated by the error bars. The curves connecting the 
points are to guide the eye. 
\end{figure}

\newpage
\begin{figure}
 \begin{center}
\epsfig{file=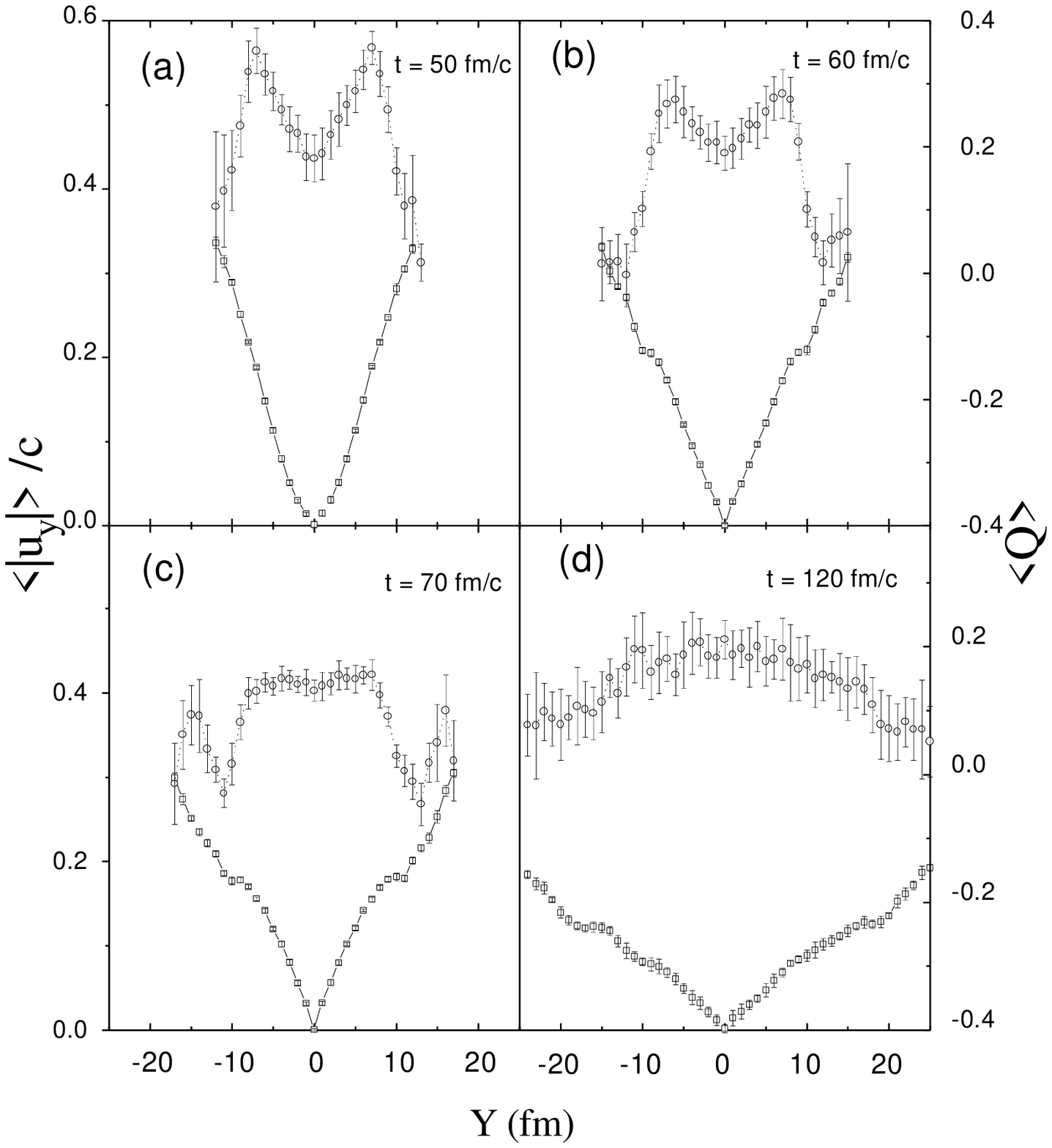,width=0.8\textwidth}
 \vspace{-0.5cm} 
 \end{center}

{\bf Fig. 3} \\
The sample averaged values of the magnitude of the y component of 
flow velocity $\langle|u_y|\rangle$ along y axis are shown by squares for 
different time steps.
The sample averaged values of the asymmetry variable $\langle Q\rangle$ along 
the y axis also shown by circles in the same graph. The sample 
fluctuations are indicated by the error bars.
The curves connecting the points are to guide the eye. 
\end{figure}
\newpage
\begin{figure}
\begin{center}
\epsfig{file=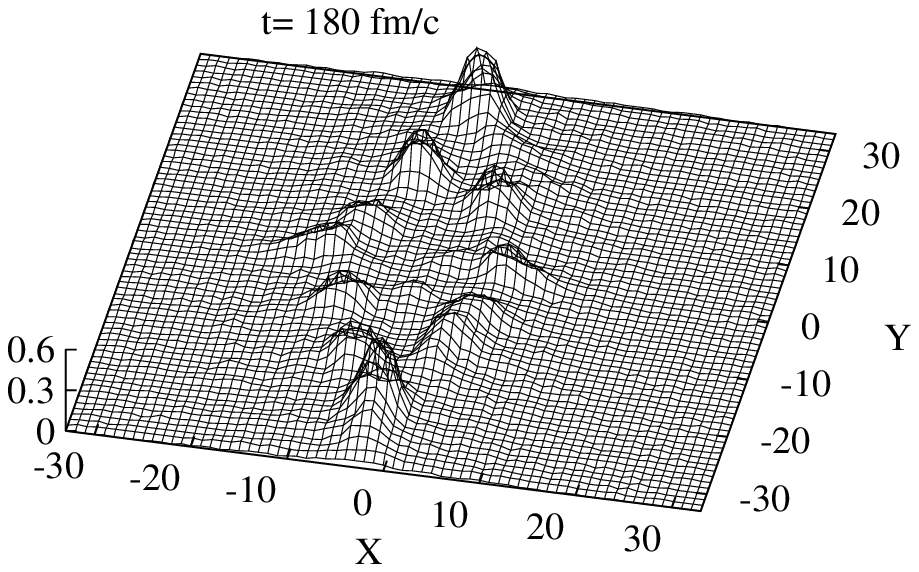,width=1.0\textwidth}
\vspace{0.5cm} 
\end{center}

{\bf Fig. 4} \\
The density profile $\rho(x,y;t)/\rho_0$ versus 
the position ($x$, $y$) for a particular sample is shown. The angles 
of view are same as that of Fig. 1.
\end{figure}

\end{document}